\documentstyle[12pt]{article}
\begin{document}
\textheight 200mm
\textwidth 160mm
\leftmargin 10mm
\newcommand{\be}{\begin{eqnarray}}
\newcommand{\ee}{\end{eqnarray}}
\newcommand{\zt}{\zeta}
\newcommand{\ve}{\epsilon}
\newcommand{\al}{\alpha}
\newcommand{\gm}{\gamma}
\newcommand{\bt}{\beta}
\newcommand{\dt}{\delta}
\newcommand{\la}{\lambda}
\newcommand{\vp}{\varphi}
\newcommand{\nn}{\nonumber}
\renewcommand{\baselinestretch}{1.4}
\newcommand{\lmx}[1]{\begin{displaymath} {#1}=
                    \left(\begin{array}{rrr}}
\newcommand{\rmx}{\end{array} \right) \end{displaymath}}
\begin{titlepage}
\begin{center}
 { \Large\bf Three--loop renormalization group analysis\\
 of a complex model with stable fixed point: \\[12pt]
 Critical exponents up to $\epsilon^3$ and $\epsilon^4$}
\end{center}
\vspace{0.5cm}
\begin{center}
 {\bf A. I. Mudrov}
\end{center}
\begin{center}
Department of Theoretical Physics,
Institute of Physics, St.Petersburg State University, Ulyanovskaya 1,
Stary Petergof, St.Petersburg, 198904, Russia\\
\end{center}
\begin{center}
{\bf K. B. Varnashev}
\end{center}
\begin{center}
Department of Physical Electronics, State Electrotechnical
University, \\
Prof. Popov Street  5, St.Petersburg, 197376, Russia
\end{center}

\begin{center}
{\bf  Abstract}
\end{center}
The complete analysis of a model with three quartic coupling
constants associated with an $O(2N)$--symmetric, a cubic, and a
tetragonal interactions is carried out  within the three--loop
approximation of the renormalization--group (RG) approach in
$D = 4 - 2\epsilon$ dimensions. Perturbation expansions for RG
functions are calculated using dimensional regularization and the
minimal subtraction (MS) scheme. It is shown that for $N\ge 2$ the
model does possess a stable fixed point in three dimensional space
of coupling constants, in accordance with predictions made earlier
on the base of the lower-order approximations. Numerical estimate for
critical (marginal) value of the order parameter dimensionality
$N_c$ is given using Pad{$\acute{\rm e}$}--Borel summation of the
corresponding $\ve$--expansion series obtained. It is observed that
two--fold degeneracy of the eigenvalue exponents in the one--loop
approximation for the unique stable fixed point leads to the
substantial decrease of the accuracy expected within three loops and
may cause powers of $\sqrt{\ve}$ to appear in the expansions.
The critical exponents $\gamma$ and $\eta$ are calculated for all
fixed points up to $\epsilon^3$ and $\epsilon^4$, respectively, and
processed by the Borel summation  method modified with a conformal
mapping. For the unique stable fixed point the magnetic
susceptibility exponent $\gamma$ for $N=2$ is found to differ
in third order in $\ve$ from that of an $O(4)$--symmetric point.
Qualitative comparison of the results given by $\epsilon$--expansion,
three--dimensional RG analysis, non--perturbative RG arguments, and
experimental data is performed.
\vspace{0.5cm}

PACS numbers: 64.60.Ak, 64.60.Fr, 75.40.Cx, 75.50.Ee

\end{titlepage}
\section{Introduction}
\label{sec:1}
There are numerous complicated models with more than two
independent quartic coupling constants. They describe phase
transitions in a variety of systems and are actively studied
within the $\epsilon$--expansion as well as by the field--theoretical
renormalization group method in three dimensions [1--10]. Critical
fluctuations in anisotropic systems with several quartic coupling
constants are known to destroy, as usual, continuous transitions
converting them into first--order ones. This fact, however,
has not been strictly proved. On the contrary, it is possible to
construct models with a large number of coupling constants whose RG
equations have stable fixed points \cite{9}. It  means that the
presence of three and more coupling constants in the Landau--Wilson
Hamiltonian does not forbid continuous phase transitions in the
system. Nevertheless, complicated models with stable fixed points
are quite rare. One of such models describing certain
antiferromagnetic phase transitions and the structural
transition in $NbO_2$ crystal will be studied in the paper.

We consider the critical behavior of a model given by the
fluctuation Landau-Wilson Hamiltonian with three quartic
interaction terms:
\begin{eqnarray}
H &=&
\int d^{~D}x \Bigl[{1 \over 2}( m_0^2 \vp^\al_i
\vp^\al_i + \partial_\mu \vp^\al_i \partial_\mu \vp^\al_i)+\nn\\
&+& {1 \over 4!} \Bigl(u_0 G^{\>\>\>\al\bt\mu\nu}_{1\> ijkl}+
v_0 G^{\>\>\>\al\bt\mu\nu}_{2\> ijkl}+ 2 z_0 G^{\>\>\>
\al\bt\mu\nu}_{3\> ijkl}\Bigr)
\vp_i^{\al} \vp_j^{\bt} \vp_k^{\mu} \vp_l^{\nu}\Bigr].
\label{eq:1.1}
\end{eqnarray}
Here $\vp_i^{\al}$, $i = 1, \ldots, N$, $\al = 1, 2$, is the real
$2N$--component order parameter field in
$D=4 - 2\ve$ dimensions and
$m_0$, $u_0$, $v_0$, $z_0$ are the bare mass and coupling constants,
respectively.
The squared bare mass $m_0^2$ can be thought of as proportional to
the deviation from the mean--field transition point (line).
The field $\vp^\al_i$ is regarded as consisting of two sets
of components, even and odd, each of them may be considered as a
real $N$--component vector.
Tensors $G_1$, $G_2$, and $G_3$ in the Hamiltonian (\ref{eq:1.1})
corresponding to isotropic, cubic, and tetragonal interactions
have the following symmetrized form:
\begin{eqnarray}
 G^{\>\>\>\al\bt\mu\nu}_{1\> ijkl}&=&\frac{1}{3}\Bigl(
\delta^{\al\bt}\delta^{\mu\nu} \delta_{ij}\delta_{kl}+
\delta^{\al\mu}\delta^{\bt\nu} \delta_{ik}\delta_{jl}+
\delta^{\al\nu}\delta^{\bt\mu} \delta_{il}\delta_{kj}\Bigr),
\nn \\
 G^{\>\>\>\al\bt\mu\nu}_{2\> ijkl}&=&
\delta^{\al\bt}\delta^{\al\mu}\delta^{\al\nu} \delta_{ij}
\delta_{ik}\delta_{il},
\label{eq:1.2} \\
 G^{\>\>\>\al\bt\mu\nu}_{3\> ijkl}&=&\frac{1}{3}\Bigl(
\delta^{1\al}\delta^{1\bt}\delta^{2\mu}\delta^{2\nu} +
\delta^{1\al}\delta^{1\mu}\delta^{2\bt}\delta^{2\nu} +
\delta^{1\al}\delta^{1\nu}\delta^{2\mu}\delta^{2\bt}
\nn \\&&
+ \delta^{2\al}\delta^{2\bt}\delta^{1\mu}\delta^{1\nu} +
\delta^{2\al}\delta^{2\mu}\delta^{1\bt}\delta^{1\nu} +
\delta^{2\al}\delta^{2\nu}\delta^{1\mu}\delta^{1\bt}
\Bigr)\delta_{ij}\delta_{ik}\delta_{il}.
\nn
\end{eqnarray}
When $u$ is equal to zero, the Hamiltonian (\ref{eq:1.1})
describes $N$ identical non--interacting anisotropic $XY$  models
\cite{10}, while for $z = 0$ it turns into the Hamiltonian of the
well--known hypercubic model.

The Hamiltonian (\ref{eq:1.1}) governs the critical thermodynamics
in a number of interesting physical systems. So, for example,
when $N = 2$ it describes the structural phase transition in
$NbO_2$ crystal and, for $v=z$, the  antiferromagnetic transitions
in $TbAu_2$ and $DyC_2$. Another physically important case $N=3$ is
relevant to the antiferromagnetic phase transition in $K_2 Ir Cl_6$
crystal and, for $v = z$, to those in $Tb D_2$ and $Nd$ \cite{1,2}.
The detailed analysis of these systems along the line of the Landau
phenomenological theory can be found in \cite{1,1a,11} with
references to the experimental works therein.

For the first time the renormalization group analysis of the
model (\ref{eq:1.1}) was performed to second order in $\epsilon$
by Mukamel and Krinsky in Refs. \cite{1,1a,2}. On this ground,
it was shown that the $2N$--component real anisotropic model
(\ref{eq:1.1}) has a unique (three--dimensionally) stable fixed
point for each $N \ge 2$. The corresponding critical exponents were
recorded and for $n=2N=4$ they were found to coincide with those of
the Heisenberg fixed point. On the other hand, the critical behavior
of this model was studied within the two--loop approximation by the
alternative RG approach in three dimensions \cite{12}.
The calculations made provided the same qualitative predictions,
although for the physically interesting cases $N=2$ and $N=3$
the critical exponents were found to be numerically close to those
of the $3D$ $XY$ model rather than the Heisenberg ones.

It is well known, however, that low--order approximations lead to
rather crude quantitative and, sometimes, contradictory qualitative
results, especially for systems with nontrivial symmetry (see, for
instance, Refs. \cite{6,7,13}).
To make more definite conclusion concerning the unique fixed point
stability and  obtain more accurate values of the critical exponents
one has to consider long enough perturbation theory series. Such
series are known to have the zero radius of convergence and therefore
are, at best, asymptotic. To extract reliable information from them a
proper resummation procedure must be applied. Recently such work for
the model under consideration was done within the field--theoretical
RG approach in three dimensions, where the three--loop expansions for
$\bt$--functions and critical exponents were calculated for arbitrary
$N$ \cite{14}. Using the generalized Pad{$\acute{\rm e}$}--Borel
resummation technique, the coordinates of all the fixed points were
found. It was shown that the unique stable fixed point did exist on
the three--dimensional RG flow diagram when $N\geq 2$.

It should be noted that, assuming $v = z$ the model (\ref{eq:1.1})
formally turns into that with generalized cubic anisotropy and the
complex order parameter field. The latter is a specific case ($m=2$)
of the well--known $mn$--component model. The critical behavior
of this model was investigated in Refs. \cite{15,16}.
Two-- and three--loop calculations done for the case $m=2$, $n\ge2$
predict stability of the mixed fixed point, the analog of the
unique stable fixed point of the model (\ref{eq:1.1}).

At the same time, there are general non--perturbative arguments
in favor of the unique stable fixed point should not be in the
physical space although its existence is not forbidden at $D>3$
\cite{17}. According to those considerations the only
three--dimensionally stable fixed point may be  the Bose one
and it is that point which governs the critical thermodynamics
in the phase transitions mentioned. The point is that the model
(\ref{eq:1.1}) describes $N$ interacting Bose systems when $v=z$.
As was shown by J. Sak \cite{18}, the interaction term can be
represented as the product of the energy
operators of various two--component subsystems. It was also found
that one of the eigenvalue exponents characterizing the evolution
of this term under the renormalization group in a neighborhood
of the Bose fixed point is proportional to the specific heat
exponent $\al$. Since $\al$ is believed to be negative at this point
(that is confirmed by highly precise up-to-date mesurements of the
specific heat exponent of liquid Helium \cite{19} including those
in outer space \cite{20} and the high--loop RG computations carried
out for the simple $O(n)$--symmetric  model in three dimensions
\cite{21,22}) the interaction is irrelevant. Therefore, the Bose
fixed point should be stable in three dimensions.

Renormalization group approach, however, when directly applied
to the model (\ref{eq:1.1}) and to the relative $mn$--component one,
has not still confirmed that non--perturbative
conclusion. On the contrary, all calculations performed up to now
indicate existence of the unique stable fixed point in the physical
space, while the Bose point appears to be three--dimensionally
unstable [1--3,14,16--18]. This may be a consequence of rather crude
approximations used, and the higher order being taken into account
the closer the perturbative results could be to the precise ones.
So, the aim of the paper is to investigate the critical behavior
of the three coupling constants model (\ref{eq:1.1}) in the next,
third order in $\ve$ and verify compatibility of predictions given
by the $\ve$--expansion method with predictions based on the other
techniques.

The main results of our study to be discussed below are as follows.
\begin{itemize}
\item The $\beta$--functions of the record length for the model
      (\ref{eq:1.1}) are obtained by the $\ve$--expansion method.
      To calculate tensor convolutions associated with the
      Feynman's graphs an algorithm was developed and a specially
      designed computer application package was written.
\item Coordinates of all fixed points and their eigenvalue
      exponents are calculated in general form within the
      three--loop approximation. The problem of stability
      of the fixed points is analyzed. The unique fixed point
      rather than the Bose one is found
      to be three--dimensionally stable in the frame of given
      approximation. Numerical estimate of the critical
      dimensionality of the order parameter $N_c$
      obtained confirms this conclusion.
\item It is observed that one--loop degeneracy of the eigenvalue
      exponents of the unique stable fixed point leads to a certain
      complication in calculating their $\ve$--expansion series.
      This problem is investigated in detail.
      It is shown that such a degeneracy substantially reduces
      the accuracy expected from given approximation and
      may result in appearance of the powers of $\sqrt{\ve}$ in
      corresponding series.
\item
      Perturbation series for the critical exponents $\gm$ and
      $\eta$ are expanded to $\ve^3$ and $\ve^4$, respectively.
      For $N=2$ the magnetic susceptibility exponent series
      of the unique stable fixed point  and the $O(4)$--symmetric
      point are found to be different
      (up to second order in $\ve$ they exactly coincide \cite{2}).
      The numerical values of the critical exponents are estimated
      by resumming the series using the Borel transformation
      with a conformal mapping.
\end{itemize}
The results of our investigation are discussed in comparison with
conclusions given by other theoretical approaches and experimental
data.

The set up of the paper is as follows. In Section \ref{sec:2} the
renormalization scheme is formulated and three--loop expansions
for the $\bt$--functions and critical exponents are presented.
Specific symmetry properties of the initial Hamiltonian
(\ref{eq:1.1}) are revealed and used as criteria of correctness of
the equations deduced. In Section \ref{sec:3} RG expansions for
coordinates of the fixed points are written out for arbitrary
$N$ and the problem of their stability is studied. The numerical
estimate of the critical dimensionality $N_c$, at which the topology
of flow diagrams changes, is obtained in  Section \ref{sec:4} by
resummation of its series using Pad{$\acute{\rm e}$}--Borel method.
The RG expansions of the critical exponents for the physically
interesting cases $N=2$, $N=3$ and their numerical estimates
are given therein. Conclusion is devoted to discussions of the
results of the investigation. The paper has two appendices.
Appendix A contains $\ve$--expansions for the eigenvalue exponents
of the fixed points for arbitrary $N$. In Appendix B we analyze
the problem of degeneracy of the eigenvalue exponents of
the unique stable fixed point and its implications.

\section{RG expansions and symmetries}
\label{sec:2}
To calculate the $\bt$--functions and critical exponents
normalizing conditions must be imposed on renormalized one--particle
irreducible inverse Green's functions $\Gamma^{(2)}_R$ and vertices
$\Gamma^{(4)}_R$ given by corresponding Feynman's diagrams. Within
the massless theory they are normalized in a  conventional way:
\begin{equation}
\begin{array}{lcrc}
\Gamma_R^{(2)}(\{p\};\mu, u, v, z)\Big\arrowvert_{
p^2 = 0
} &=& 0 &,
\nonumber \\
{\partial \over {\partial p^2}} \Gamma_R^{(2)} (\{p\}; \mu, u, v, z)
\Big\arrowvert_{p^2 = \mu^2} &=& 1 &,
\nonumber \\
\Gamma_{UR}^{(4)} (\{p\}; \mu, u, v, z)&=& \mu^{\ve} u &,
\label{eq:2.1} \\
\Gamma_{VR}^{(4)} (\{p\}; \mu, u, v, z)&=&\mu^{\ve} v &,
\nonumber \\
\Gamma_{ZR}^{(4)} (\{p\}; \mu, u, v, z)&=&\mu^{\ve} z &,
\nonumber
\end{array}
\end{equation}
with one more condition on the $\vp^2$ insertion
\begin{eqnarray}
\Gamma_R^{(1,2)}(\{p\},\{q\};\mu,u,v,z)
\Big\arrowvert_{{p^2 = q^2  = \mu^2 }\atop{pq=-{1\over3}}\mu^2}
&=& 1
\ \ . \label{eq:2.2}
\end{eqnarray}
Here $m$, $u$, $v$, and $z$ are the renormalized mass and
dimensionless coupling constants, with $\mu$ being an arbitrary
mass parameter introduced for dimensional regularization.

The vertices $\Gamma_u^{(4)}$, $\Gamma_v^{(4)}$, $\Gamma_z^{(4)}$
are connected with the vertex function without external lines
normalized in the following way:
$$
 \Gamma_{\>\>\>ijkl}^{(4)\>\>\>\al\bt\mu\nu}=
 \Gamma_u^{(4)} \cdot G^{\>\>\>\al\bt\mu\nu}_{1\> ijkl}+
 \Gamma_v^{(4)} \cdot G^{\>\>\>\al\bt\mu\nu}_{2\> ijkl}+
 \Gamma_z^{(4)} \cdot G^{\>\>\>\al\bt\mu\nu}_{3\> ijkl}.
$$
From renormalization conditions (\ref{eq:2.1}) and (\ref{eq:2.2})
the expansions for the renormalization constants $Z_{\vp}$,
$Z_u$, $Z_v$, $Z_z$, and $Z_{\vp^2}$ may be obtained.
These constants relate the bare mass $m_0$ and three coupling
constants $u_0$, $v_0$, $z_0$ of the Hamiltonian (\ref{eq:1.1}) to
the corresponding physical parameters:
\begin{equation}
   m_0^2 = \frac{Z_{\vp^2}}{Z_{\vp}} m^2=Z_m m^2,\quad
   u_0   = \mu^{\ve} \frac{Z_u}{Z^2_{\vp}} u, \quad
   v_0   = \mu^{\ve} \frac{Z_v}{Z^2_{\vp}} v, \quad
   z_0   = \mu^{\ve} \frac{Z_z}{Z^2_{\vp}} z. \quad
\label{eq:2.3}
\end{equation}
Thus, with relations (\ref{eq:2.3}) taken  into account, the
$\bt$--functions and critical exponents can be calculated via the
formulas
\begin{equation}
\begin{array}{lcr}
\frac{\partial \ln u_0}{\partial u} \bt_u +
\frac{\partial \ln u_0}{\partial v} \bt_v +
\frac{\partial \ln u_0}{\partial z} \bt_z &=& - \ve,
\nn \\
\frac{\partial \ln v_0}{\partial u} \bt_u +
\frac{\partial \ln v_0}{\partial v} \bt_v +
\frac{\partial \ln v_0}{\partial z} \bt_z &=& - \ve,
\label{eq:2.4} \\
\frac{\partial \ln z_0}{\partial u} \bt_u +
\frac{\partial \ln z_0}{\partial v} \bt_v +
\frac{\partial \ln z_0}{\partial z} \bt_z &=& - \ve,
\nn
\end{array}
\end{equation}
\begin{equation}
\begin{array}{lcr}
\eta(u,v,z)&=&
2 \frac{\partial \ln Z_{\vp}}{\partial u} \bt_u +
2 \frac{\partial \ln Z_{\vp}}{\partial v} \bt_v +
2 \frac{\partial \ln Z_{\vp}}{\partial z} \bt_z,
\nn \\
\eta_2(u,v,z)&=& \>\>\>\>
2 \frac{\partial \ln Z_{\vp^2}}{\partial u} \bt_u +
2 \frac{\partial \ln Z_{\vp^2}}{\partial v} \bt_v +
2 \frac{\partial \ln Z_{\vp^2}}{\partial z} \bt_z ,
\label{eq:2.5}
\end{array}
\end{equation}
where $\bt_g \equiv \frac{\partial g}{\partial |\ln \mu|}$,
$g=u,v,z$. The critical exponents $\eta$ and $\eta_2$ are found
by substituting zeros of the $\bt$--functions into expressions
(\ref{eq:2.5}). The critical exponent $\gm$ is given by the well
known scaling relation $\gm^{-1} = 1 + \frac {\eta_2}{2 - \eta}$.

The contribution of a Feynman's graph into an RG--function comprises
three factors: the combinatorial coefficient, the result of
tensor convolution and the integral value associated to the diagram.
The combinatorial factors, and the values of integrals are known
from Ref. \cite{23}, while evaluating tensor convolutions for vertex
and mass diagrams is the problem to be solved. To do it we have
developed a computer  application package written in PASCAL. The
algorithm is based upon two quite natural assumptions:
\begin{enumerate}
\item Tensor convolution algebra is closed, i.e.
   each monomial $G_{i_1} * \ldots * G_{i_{l+1}}$ contributing to a
   vertex function is a linear combination of the basic tensors
   $G_1$, $G_2$, and $G_3$:
\be
  G_{i_1} * \ldots * G_{i_{l+1}}&=&a(N)G_1+ b(N)G_2+ c(N)G_3.
  \label{Conv}
\ee
\item Dependence of the coefficients $a(N)$, $b(N)$, and $c(N)$ upon
      $N$ is of polynomial character. The degree of the polynomials
      does not exceed the number $l$ of loops in the Feynman's graph.
\end{enumerate}
The first condition means that one has no new interactions generated
in the model (\ref{eq:1.1}). The second proposition becomes evident
upon analyzing the particular form of the tensors $G_i$.

Since a polynomial of degree $l$ is fully determined by its values
in $l+1$ different points, it is sufficient to compute convolutions
consecutively assuming $N=2,\ldots l+2$ (the reason to start from
$2$ is linear dependence between $G_i$ when $N=1$). In order to
evaluate three indeterminates $a(N)$, $b(N)$, and $c(N)$ we
compare both sides of expression (\ref{Conv}), having assigned
values $(^{1122}_{1122})$, $(^{1111}_{1111})$, and
$(^{1212}_{1111})$ to the multi--index $(^{\al\bt\mu\nu}_{ijkl})$.
It provides a non--degenerate system of linear equations whose
$3\times 3$--matrix does not depend on $N$.
From this system the coefficients of decomposition (\ref{Conv}) are
found. Similar procedure was applied to the mass diagrams. The
results of our computations recover those achieved within the
four--loop approximation for simple $O(n)$--symmetric model
\cite{23}.

After some work, we obtain the expressions for the RG--functions
within the three--loop approximation (Fisher's exponent $\eta$ is
calculated up to four loops) using dimensional regularization
\cite{24} and the MS scheme \cite{25}:
\begin{displaymath}
\begin{array}{lll}
\bt_u &=& \ve u-u^2 - \frac{1}{2 (N + 4)} \Bigl(6 u v + 2 u z
 \Bigr) +
\\&&
\frac{1}{4 (N + 4)^2} \Bigl[12 u^3 (3 N + 7) + 132 u^2 v +
44 u^2 z + 30 u v^2 + 10 u z^2 \Bigr] -
\\&&
\frac{1}{16 (N + 4)^3}
\Bigl[4 u^4 (48 \zt(3) (5 N + 11) + 33 N^2 + 461 N + 740) +
\\&&
12 u^3 v (384 \zt(3) + 79 N + 659) +
4 u^3 z ( 384 \zt(3) + 79 N + 659) +
\\&&
18 u^2 v^2 (96 \zt(3) + N + 321) + 1380 u^2 v z +
2 u^2 z^2 (288 \zt(3) + 3 N +
\\&&
733) + 1512  u v^3 + 18 u v^2 z +
504 u v z^2 + 222 u z^3 \Bigr],
\end{array}
\end{displaymath}

\begin{equation}
\begin{array}{lll}
\bt_v &=& \ve v-\frac{1}{2 (N + 4)} (12 u v + 9 v^2 + z^2) +
\\&&
\frac{1}{4 (N + 4)^2} \Bigl[4 u^2 v (5 N + 41) + 276 u v^2 +
20 u v z + 24 u z^2 + 102 v^3 + 10 v z^2 +
\\&&
8 z^3 \Bigr]-
\frac{1}{16 (N + 4)^3} \Bigl[8 u^3 v (96 \zt(3) (N + 7)-
13 N^2 + 184 N + 821) +
\\&&
18 u^2 v^2 (768 \zt(3) + 17 N + 975) +
12 u^2 v z (96 \zt(3) - 13 N + 154) +
\\&&
2 u^2 z^2 (576 \zt(3) + 43 N + 667) +
108 u v^3 (96 \zt(3) + 131) + 306 u v^2 z +
\\&&
12 u v z^2 (96 \zt(3) + 187) +
2 u z^3 (384 \zt(3) + 395) +
27 v^4 (96 \zt(3) + 145) +
\\&&
162 v^2 z^2 +
8 v z^3 (48 \zt(3) + 101) +
3 z^4 (32 \zt(3) + 17) \Bigr], \label{eq:2.6}
\end{array}
\end{equation}

\be
\bt_z &=& \ve z-\frac{1}{2 (N + 4)} (12 u z + 6 v z + 4 z^2) +
\nn\\&&
\frac{1}{4 (N + 4)^2} \Bigl[4 u^2 z (5 N + 41) + 204 u v z
+ 116 u z^2 + 30 v^2 z + 72 v z^2 + 18 z^3 \Bigr]-
\nn\\&&
\frac{1}{16 (N + 4)^3} \Bigl[8 u^3 z (96 \zt(3) (N + 7)
-13 N^2 + 184 N + 821) +
\nn\\&&
12 u^2 v z (864 \zt(3) + 4 N + 1129) +
4 u^2 z^2 (1440 \zt(3) + 47 N + 1796) +
\nn\\&&
18 u v^2 z (192 \zt(3) + 391) + 72 u v z^2 (96 \zt(3) + 103) +
2 u z^3 (960 \zt(3) + 1517) +
\nn\\&&
1512 v^3 z +
36 v^2 z^2 (48 \zt(3) + 35) +
72 v z^3 (16 \zt(3) + 25) +
4 z^4 (48 \zt(3) + 91) \Bigr],
\nn\\[12pt]
\gamma^{-1} &=&
1-\frac{1}{2(N+4)} (2 u (N+1)+3 v+z)+
\nn\\&&
\frac{1}{2(N+4)^2}\Bigl[6 u^2 (N+1)+18 u v+6 u z+9 v^2
+3 z^2\Bigr]-
\nn\\&&\frac{1}{16(N+4)^3}
\Bigl[12 u^3 (N+1) (11 N+39)+54 u^2 v (11 N+39)+
\label{eq:2.7}\\&&
18 u^2 z (11 N+39)+6 u v^2 (5 N+398)+564 u v z+
\nn\\&&
2 u z^2 (5 N+304)+801 v^3+15 v^2 z+267 v z^2+117 z^3\Bigr],
\nn
\ee

\be
\eta &=&\frac{1}{2(N+4)^2}
\Bigl(2 u^2 (N+1)+6 u v+2 u z+3 v^2+z^2\Bigr)-
\nn\\&&\frac{1}{8(N+4)^3}
\Bigl[4 u^3 (N+1) (N+4)+18 u^2 v (N+4)+6 u^2 z (N+4)+
\nn\\&&
81 u v^2+18 u v z+21 u z^2+27 v^3+9 v z^2+4 z^3\Bigr]-
\label{eq:2.8}\\&&\frac{1}{32(N+4)^4}
\Bigl[40 u^4 (N+1) (N^2-9 N-25)+240 u^3 v (N^2-9 N-25)+
\nn\\&&
80 u^3 z (N^2-9 N-25)-180 u^2 v^2 (N+58)+360 u^2 v z (N-8)-
\nn\\&&
180 u^2 z^2 (N+14)-7020 u v^3-180 u v^2 z-2340 u v z^2-
\nn\\&&
1020 u z^3-1755 v^4-90 v^2 z^2-720 v z^3-75 z^4\Bigr],
\nn
\ee
where $\zt$ is the Riemann $\zt$--function: $\zt(3)=1.20206$.
Expressions (\ref{eq:2.6}) -- (\ref{eq:2.8}) are in accordance
with those obtained earlier in Ref. \cite{2}, where corresponding
calculations for RG functions were carried out to $\ve^2$, and,
assuming $v=z\equiv 0$ and $N={n\over 2}$, with results of
Ref. \cite{26}, in which the critical exponents of the well--known
$O(n)$--symmetric model  were found up to $\ve^4$.
If $u\equiv 0$ and $v=z$, the right--hand side of the second (third)
equation (\ref{eq:2.6}) goes over into the $\bt$--function for
Bose--like systems, the coupling constants being normalized properly.
The latter, obviously,  coincides with that of the $O(n)$--symmetric
model when $n=2N=2$.

In conclusion of this section, let us formulate a criterion to
check the correctness of the results obtained. It relies on a
specific symmetry property of the Hamiltonian (\ref{eq:1.1}) of the
system under consideration \cite{12}. It occurs that transformation
\begin{equation}
\begin{array}{lcl}
\varphi_{2N-1} &\rightarrow& {1\over {\sqrt{2}}}~(\varphi_{2N-1}
+ \varphi_{2N}) , \\
&& \\
\varphi_{2N} &\to& {1\over {\sqrt{2}}}~(\varphi_{2N-1} -
\varphi_{2N}),
\label{eq:2.9}
\end{array}
\end{equation}
combined with substitution of quartic coupling constants
\begin{equation}
u \to u, \quad v \to  {1\over 2}(v + z), \quad
z \to  {1\over 2}(3 v - z)
\label{eq:2.10}
\end{equation}
does not change the structure of the Hamiltonian itself.

Similar situation takes place for $N = 1$ and $z = 0$
in the case of another field transformation
\begin{equation}
\begin{array}{lcl}
\varphi_1 &\rightarrow& {1\over {\sqrt{2}}}~(\varphi_1
+ \varphi_2) , \\
&& \\
\varphi_2 &\rightarrow& {1\over {\sqrt{2}}}~(\varphi_1 -
\varphi_2) ,
\label{eq:2.11}
\end{array}
\end{equation}
which does not affect the Hamiltonian resulting only
in the following replacement of $u$ and $v$:
\begin{equation}
u \rightarrow u + {3 \over 2}~v, \quad v \rightarrow - v .
\label{eq:2.12}
\end{equation}

It is well known that the RG equations should be invariant with
respect to any transformation conserving the structure of the
Hamiltonian \cite{27}. It means that for every $N$, in the case
of symmetry (\ref{eq:2.10}), functions $\beta_u$, $\beta_v$,
and $\beta_z$ should obey special relations which may be
readily written down:
\begin{equation}
\begin{array}{rcr}
\beta_u ( u,~ v,~ z ) &=& \beta_u
\biggl( u,~ {1 \over 2}~(v + z),
~{1 \over 2}~(3 v - z) \biggr) \ \ ,\\
\beta_v ( u,~ v,~ z ) + \beta_z ( u,~ v,~ z ) &=&
2\beta_v  \biggl( u,~ {1 \over 2}~(v + z),~ {1
\over 2}~(3 v - z)
\biggr) \ \ ,\\
3 \beta_v ( u,~ v,~ z ) - \beta_z ( u,~ v,~ z ) &=&
2\beta_z \biggl( u,~ {1 \over 2}~(v + z),~ {1
\over 2}~(3 v - z)
\biggr) \ \ .
\end{array}\label{eq:2.13}
\end{equation}
For $N = 1$ and $z = 0$ the other symmetry (\ref{eq:2.12}) results
in
\begin{equation}
\begin{array}{rcr}
\beta_u ( u,~ v,~ 0 ) + {3 \over 2}~ \beta_v ( u,~ v,~ 0 ) &=&
\beta_u \biggl( u + {3 \over 2}~v,~ - v,~ 0 \biggr) \ \ , \\
\beta_v ( u,~ v,~ 0 ) &=& - \beta_v \biggl( u + {3
\over 2}~v,~ - v,~ 0
\biggr) \ \ .
\end{array}\label{eq:2.14}
\end{equation}
At last, the critical exponents are invariant
under the transformations (\ref{eq:2.10}) and (\ref{eq:2.12}).
So, the first symmetry
gives
\begin{equation}
\begin{array}{rcr}
\gm^{-1} ( u,~ v,~ z ) &=&
\gm^{-1} \biggl( u,~ {1 \over 2}~(v + z),
~{1 \over 2}~(3 v - z) \biggr) \ \ ,\\
\eta ( u,~ v,~ z ) &=&
\eta \biggl( u,~ {1 \over 2}~(v + z),
~{1 \over 2}~(3 v - z) \biggr) \ \ .
\end{array}\label{eq:2.15}
\end{equation}
Similar relations should take place in the case of the symmetry
(\ref{eq:2.12}). It can be  easily verified that conditions
(\ref{eq:2.13})--(\ref{eq:2.15}) are satisfied indeed.

Symmetries of the initial Hamiltonian like those described above
(such symmetries do not exist always and to find them requires
certain efforts) play, in some cases, an extremely important role.
Namely, the series being obtained within the  field--theoretical
RG approach in $3D$ are necessarily processed with the use of some
resummation procedure (e.g. Pad{$\acute{\rm e}$},
Pad{$\acute{\rm e}$--Borel,
Chisholm--Borel etc.), and satisfaction of the numerical results to
the exact symmetry relations serves as a criterion to estimate the
accuracy expected from the approximation scheme employed \cite{6}.

\section{Fixed points and stability}
\label{sec:3}

Two critical exponents $\gm$ and $\eta$ are known to completely
specify the critical behavior of a system \cite{28}. They are
determined from RG functions by going to the infrared--stable fixed
points $g_c=(u_c, v_c, z_c)$, which are found as zeros of the
$\bt$--functions in the form of series in powers of $\ve$:
$$g_c=g_c(\ve)=\sum^\infty_{k=1}g_k \ve^k .$$
There exist eight fixed points in the model under consideration
\cite{2,12}, one of them (Gaussian) is trivial:\\[12pt]
\begin{tabular}{llll}
{\bf 1.}&\multicolumn{3}{l}{\bf  Gaussian fixed point}
\\[6pt]
&$u_c$&=&$v_c\>=\>z_c\>=\>0.$
\\[6pt]
{\bf 2.}&\multicolumn{3}{l}{\bf $O(2N)$--symmetric or Heisenberg
fixed point}
\\[6pt]
&$u_c$&=&$ \ve+\frac{3 (3 N+7)}{(N+4)^2} \ve^2-
\Bigl(\frac{12 \zt(3) (5 N+11)}{(N+4)^3}+\frac{33 N^3-55 N^2-440 N
-568}{4 (N+4)^4}\Bigr)
\ve^3$,
\\
&$v_c$&=&$z_c\>=\>0$.
\\[6pt]
{\bf 3.}&\multicolumn{3}{l}{\bf Ising fixed point}
\\[6pt]
&$u_c$&=&$z_c\>=\>0$,
\\
&$v_c$&=&$\frac{2(N+4)}{9}\ve + \frac{68(N+4)}{243} \ve^2+
\Bigl(\frac{709 (N+4)}{6561}-\frac{32 (N+4)}{81}\zt(3) \Bigr)\ve^3$.
\\[6pt]
{\bf 4.}&\multicolumn{3}{l}{\bf Cubic fixed point}
\\[6pt]
&$u_c$&=&$\frac{N+4}{3 N}\ve+\frac{N+4}{81 N^3} (1-2 N) (19 N-53)
 \ve^2+$
\\[6pt]&&
&$\Bigl(\frac{4(N+4)}{27 N^4} \zt(3)  (8 N^3-12 N^2-7 N+7)-$
\\[6pt]&&
&$\frac{N+4}{8748 N^5} (3910 N^4+41971 N^3-114987 N^2+90160 N
-22472)\Bigr) \ve^3$,
\\[6pt]
&$v_c$&=&
$\frac{2(N+4)}{9 N} (N-2)  \ve+\frac{2(N+4)}{243 N^3}  (2 N-1)
 (17 N^2+55 N-106) \ve^2+$
\\[6pt]&&
&$\Bigl(-\frac{16(N+4)}{81 N^4} \zt(3) (2 N^4+8 N^3-10 N^2-9 N+7)+$
\\[6pt]&&
&$\frac{N+4}{13122 N^5} (1418 N^5+11713 N^4+90281 N^3-247414 N^2
+187528 N-$
\\[6pt]&&
&$44944)\Bigr) \ve^3$,
\\
&$z_c$&=&$0$.
\\[6pt]
{\bf 5.}&\multicolumn{3}{l}{\bf Bose fixed point}
\\[6pt]
&$u_c$&=&$0$,
\\
&$v_c$&=&
$\frac{N+4}{5}\ve+\frac{6(N+4)}{25}\ve^2+\frac{N+4}{1250}(103
-384 \zt(3))\ve^3$,
\\
&$z_c$&=&
$\frac{N+4}{5}\ve+\frac{6(N+4)}{25}\ve^2+\frac{N+4}{1250}
(103-384 \zt(3))\ve^3$.
\\[6pt]
{\bf 6.}&\multicolumn{3}{l}{\bf VZ--cubic fixed point}
\\[6pt]
&$u_c$&=&$0$,
\\
&$v_c$&=&
$\frac{N+4}{9}\ve+\frac{34(N+4)}{243}\ve^2+\frac{N+4}{13122}
(709-2592\zt(3))
\ve^3$,
\\
&$z_c$&=&
$\frac{N+4}{3}\ve+\frac{34(N+4)}{81}\ve^2+\frac{N+4}{4374}
(709-2592 \zt(3))
\ve^3$.
\\[6pt]
\end{tabular}

\begin{tabular}{llll}
{\bf 7.}&\multicolumn{3}{l}{\bf I-tetragonal fixed point}
\\[6pt]
&$u_c$&=&
$\frac{N+4}{3 N}\ve+\frac{N+4}{81 N^3}(1-2 N)(19 N-53)\ve^2+$
\\[6pt]&&
&$\Bigl(\frac{4(N+4)}{27 N^4}\zt(3)(8 N^3-12 N^2-7 N+7)-$
\\[6pt]&&
&$\frac{N+4}{8748 N^5}(3910 N^4+41971 N^3-114987 N^2+90160 N
-22472)\Bigr)\ve^3$,
\\[6pt]
&$v_c$&=&
$\frac{N+4}{9 N}(N-2)\ve+ \frac{N+4}{243N^3}(2N-1)(17N^2+55N
-106)\ve^2-$
\\[6pt]&&
&$\Bigl(\frac{8(N+4)}{81N^4}\zt(3)(2N^4+8N^3 -10N^2-9N+7)-$
$\frac{N+4}{26244 N^5}(1418 N^5+$
\\[6pt]&&
&$11713N^4+90281N^3-247414N^2+187528N-44944)\Bigr)\ve^3$,
\\[6pt]
&$z_c$&=&
$\frac{N+4}{3N}(N-2)\ve+\frac{N+4}{81N^3}(2N-1)(17N^2+55N
-106)\ve^2-$
\\[6pt]&&
&$\Bigl(\frac{8(N+4)}{27N^4} \zt(3)(2N^4+8N^3-10N^2-9N+7)
-\frac{N+4}{8748N^5}
(1418N^5+$
\\[6pt]&&
&$11713N^4+90281N^3-247414N^2+187528N-44944)\Bigr)\ve^3$.
\\[6pt]
{\bf 8.}&\multicolumn{3}{l}{\bf II-tetragonal fixed point}
\\[6pt]
&$u_c$&=&
$\frac{N+4}{(5N-4)}\ve+
\frac{N+4}{(4-5 N)^3}(70 N^2-205 N+139)\ve^2+$
\\[6pt]&&
&$\Bigl(\frac{12(N+4)}{(5 N-4)^4}\zt(3)(64N^3-188N^2+151N-23)+$
\\[6pt]&&
&$\frac{N+4}{4(4-5N)^5}(6370N^4+24149N^3-144719N^2+197208N
-83256)\Bigr)\ve^3$,
\\[6pt]
&$v_c$&=&
$\frac{N+4}{(5 N-4)}(N-2)\ve+
\frac{N+4}{(5 N-4)^3}(30 N^3+25 N^2-217 N+166)\ve^2-$
\\[6pt]&&
&$\Bigl(\frac{24(N+4)}{(5 N-4)^4} \zt(3)(8N^4+16N^3-88N^2+75N-9)
-\frac{N+4}{4 (5 N-4)^5}(1030 N^5+$
\\[6pt]&&
&$2751 N^4+46033 N^3-207590 N^2+267336 N-109808)
\Bigr)\ve^3$,
\\[6pt]
&$z_c$&=&
$\frac{N+4}{5 N-4} (N-2) \ve+
\frac{N+4}{(5 N-4)^3}(30 N^3+25 N^2-217 N+166) \ve^2-$
\\[6pt]&&
&$\Bigl(\frac{24(N+4)}{(5 N-4)^4}\zt(3) (8 N^4+16 N^3-88 N^2
+75 N-9)-\frac{N+4}{4 (5 N-4)^5} (1030 N^5+$
\\[6pt]&&
&$2751 N^4+46033 N^3-207590 N^2+267336 N-109808)\Bigr)\ve^3$.
\\[6pt]&&&
\end{tabular}

From these expressions it is seen that for the physically interesting
case $N=2$ the coordinates of the fixed points 2, 4, 7, and 8
coincide in the one--loop approximation, i.e. the Heisenberg point
$u_c=\ve$, $v_c=z_c=0$ is four--fold degenerate. Such strong
degeneracy is occasional, however, and lifted out in higher orders
of the perturbation theory. So, the two--loop approximation splits
those points apart. This situation is typical for a number of
complicated models (see, for example, Refs. \cite{6,8}).

One can also notice from the above list that the Heisenberg and the
Ising fixed points coincide at $N={1 \over 2}$, while for the
components of the cubic and the Ising fixed points the relation
$-v^c_c = v^I_c$ holds at $N=1$. With $N$ increasing, the cubic
fixed point approaches the Heisenberg point from below and crosses
it at $N=N_c$, changing the sign of its $v$--coordinate. Further,
when $N\to \infty$ it moves towards the Ising point. Note, that the
II--tetragonal fixed point is getting close to the Bose one when
$N$ grows. Such a behavior of the fixed points is in accordance
with results obtained within the RG analysis in three dimensions
\cite{12,14}.

Since the symmetry transformations (\ref{eq:2.10}), (\ref{eq:2.12})
do not affect the form of the RG equations, they can only rearrange
the fixed points. This observation may be used as an additional
criterion for verification of our results. For example,
points 1, 2, 5, and 8 stay untouched under transformation
(\ref{eq:2.10}), while points 3 and 4 turn into 6 and 7,
respectively (and {\em vice versa}).

Now let us discuss the character of the stability of the fixed
points found. It is known to be determined by the signs of the
eigenvalues $\la_1$, $\la_2$, and $\la_3$ of the matrix
\lmx{M_{ij}}
    \frac{\partial\bt_u}{\partial u} &
    \frac{\partial\bt_u}{\partial v}
  & \frac{\partial\bt_u}{\partial z} \\
    \frac{\partial\bt_v}{\partial u}
  & \frac{\partial\bt_v}{\partial v}
  & \frac{\partial\bt_v}{\partial z} \\
    \frac{\partial\bt_z}{\partial u}
  & \frac{\partial\bt_z}{\partial v}
  & \frac{\partial\bt_z}{\partial z}
\rmx
evaluated at $u=u_c$, $v=v_c$, and $z=z_c$.
If the real parts of all the eigenvalue exponents are negative,
the corresponding fixed point is infrared stable in three
dimensional $(u,v,z)$--space. Besides, the "saddle--knot" type
fixed points may occur on the phase diagram, provided their
eigenvalue exponents are of opposite signs. General expressions
for the eigenvalue exponents are written out for arbitrary $N$
in Appendix A. For the interesting cases $N=2$ and $N=3$ relevant
to the substances of concern they are presented in Table 1 and
Table 2, where $\ve={1\over 2}$ corresponds to the  physical
case.

It is seen from the tables that the Ising point has single
negative eigenvalue, therefore it is stable only on the $v$--axis.
The Heisenberg point is stable on the axis too  if $N>N_c$, becoming
stable within the plane $(u,v)$ for $N<N_c$. The cubic fixed point
has the critical behavior opposite to that of the Heisenberg one;
they interchange their stability at $N=N_c$. The Bose point is
stable within the plane $(v,z)$, being of the "saddle--knot" type
in the three--parameter space. Note that the eigenvalue exponents
of points 3 and 6 are the same as well as those of  4 and 7. It
is a consequence of the symmetry (\ref{eq:2.10}) of the initial
Hamiltonian.

\begin{table}
\caption{Eigenvalue exponents for $N=2$ to third order in $\ve$.}
\vspace{0.5cm}
\hspace{2cm}
\begin{tabular}{|c|l|l|}\hline
No &Type of fixed point &\multicolumn{1}{|c|}{Eigenvalues}
\\  \hline
   &          &\\
1  & Gaussian  &$\la_u=\la_v=\la_z=\ve$
                   \\[6pt] \hline
   &          &\\
2  & Heisenberg&$\la_u=-\ve +{13 \over 12}\ve^2
                 -\frac{84\zt(3)+65}{36}\ve^3$\\
   &           &$\la_v=\la_z= {1 \over 3}\ve^2
                +\frac{1-5 \zt(3)}{6}\ve^3$
                   \\[6pt] \hline
&&\\
3  & Ising     &$\la_u=\la_z={1 \over 3}\ve
   -{38 \over 81}\ve^2 +\frac{2592\zt(3)-937}{2187}\ve^3$\\
   &           &$\la_v=-\ve+{34 \over 27}\ve^2
   -\frac{2592\zt(3)+1603}{729}\ve^3$
                   \\[6pt] \hline
&&\\
   &           &$\la_1=-\ve +{13 \over 12}\ve^2
   -\frac{84\zt(3)+65}{36}\ve^3$\\
4  & Cubic     &$\la_2=-{1 \over 3}\ve^2
   +\frac{15\zt(3)+1}{18}\ve^3$\\
   &           &$\la_z= {1 \over 3}\ve^2
   -\frac{15\zt(3)-7}{18}\ve^3$
                   \\[6pt] \hline
&&\\
   &           &$\la_u= {1 \over 5}\ve
   -{14 \over 25}\ve^2 +\frac{768\zt(3)-311}{625}\ve^3$\\
5  & Bose      &$\la_1=-{1 \over 5}\ve
   +{2 \over 5}\ve^2 -\frac{768 \zt(3)+29}{625}\ve^3$\\
   &           &$\la_2=- \ve +{6 \over 5}\ve^2
   -\frac{384\zt(3)+257}{125}\ve^3$
                   \\[6pt] \hline
&& \\
6  &$VZ$--cubic &$\la_u=\la_1={1 \over 3}\ve
   - {38 \over 81}\ve^2 +\frac{2592\zt(3)-937}{2187}\ve^3$\\
   &            &$\la_2=-\ve+{34 \over 27}\ve^2
   -\frac{2592\zt(3)+1603}{729}\ve^3$
                   \\[6pt] \hline
&&\\
   &            &$\la_1={1 \over 3}\ve^2
   -\frac{15\zt(3)-7}{18}\ve^3$\\
7  &I-tetragonal&$\la_2=-{1 \over 3}\ve^2
   +\frac{15\zt(3)+1}{18}\ve^3$\\
   &            &$\la_3= -\ve +{13 \over 12}\ve^2
   -\frac{84\zt(3)+65}{36}\ve^3$
                   \\[6pt] \hline
&&\\
   &             &$\la_1=\la_2=-{1 \over 3}\ve^2$\\
8  &II--tetragonal&$\la_3=-\ve +{13 \over 12}\ve^2
  -\frac{84\zt(3)+65}{36}\ve^3$
\\[6pt]
\hline
\end{tabular}
\end{table}

\newpage
\begin{table}
\caption{Eigenvalue exponents for $N=3$ to third order in $\ve$.}
\vspace{0.5cm}
\hspace{2cm}
\begin{tabular}{|c|l|l|}\hline
No &Type of fixed point &\multicolumn{1}{|c|}{Eigenvalues}
\\  \hline
   &          &\\
1  & Gaussian  &$\la_u=\la_v=\la_z=\ve$
                   \\[6pt] \hline
&&\\
2  & Heisenberg&$\la_u=-\ve +{48 \over 49}\ve^2
   -\frac{2(2184\zt(3)+1931)}{2401}\ve^3$\\
   &           &$\la_v=\la_z={1 \over 7}\ve
   + {104 \over 343}\ve^2 -\frac{2(5208\zt(3)-2311)}{16807}\ve^3$
                   \\[6pt] \hline
&&\\
3  & Ising     &$\la_u=\la_z={1 \over 3}\ve
   -{38 \over 81}\ve^2 +\frac{2592\zt(3)-937}{2187}\ve^3$\\
   &           &$\la_v=-\ve+{34 \over 27}\ve^2
   -\frac{2592\zt(3)+1603}{729}\ve^3$
                   \\[6pt] \hline
&&\\
   &           &$\la_1=-\ve +{250\over 243 }\ve^2
   -\frac{246888\zt(3)+165287}{118098}\ve^3$\\
4  & Cubic     &$\la_2=-{1 \over 9}\ve
  -{250 \over 2187}\ve^2 +\frac{5(36936\zt(3)
  +41611)}{1062882}\ve^3$\\
   &           &$\la_z={1 \over 9}\ve + {520 \over 2187}\ve^2
   -\frac{2(120528\zt(3)-16033)}{531441}\ve^3$
                   \\[6pt] \hline
&&\\
   &           &$\la_u= {1 \over 5}\ve -{14 \over 25}\ve^2
   +\frac{768\zt(3)-311}{625}\ve^3$\\
5  & Bose      &$\la_1=-{1 \over 5}\ve +{2 \over 5}\ve^2
   -\frac{768 \zt(3)+29}{625}\ve^3$\\
   &           &$\la_2=- \ve +{6 \over 5}\ve^2
   -\frac{384\zt(3)+257}{125}\ve^3$
                   \\[6pt] \hline
&&\\
6  &$VZ$--cubic&$\la_u=\la_1={1 \over 3}\ve
  - {38 \over 81}\ve^2 +\frac{2592\zt(3)-937}{2187}\ve^3$\\
   &            &$\la_2=-\ve+{34 \over 27}\ve^2
   -\frac{2592\zt(3)+1603}{729}\ve^3$
                   \\[6pt] \hline
&&\\
   &            &$\la_1={1 \over 9}\ve
   + {520 \over 2187}\ve^2 -\frac{2(120528\zt(3)
   -16033)}{531441}\ve^3$\\
7  &I-tetragonal&$\la_2=-{1 \over 9}\ve
  -{250 \over 2187}\ve^2 +\frac{5(36936\zt(3)
  +41611)}{1062882}\ve^3$\\
   &            &$\la_3=-\ve +{250\over 243 }\ve^2
   -\frac{246888\zt(3)+165287}{118098}\ve^3$
                   \\[6pt] \hline
&&\\
   &             &$\la_1=-{1 \over 11}\ve - {2 \over 11}\ve^2$\\
8  &II--tetragonal&$\la_2=-{1 \over 11}\ve +{2 \over 605}\ve^2$\\
   &             &$\la_3=-\ve +{58\over 55 }\ve^2
   -\frac{3(123600\zt(3)+71621)}{166375}\ve^3$
                   \\[6pt] \hline
\end{tabular}
\end{table}

The most intriguing is the II--tetragonal fixed point proving to be
absolutely stable in $3D$, as it follows from the tables. Obviously,
simple resummation procedures, such as Pad{$\acute{\rm e}$} and
Pad{$\acute{\rm e}$}--Borel methods, applied to $\la$'s do not
dismiss this conclusion. The presence of such a stable point is
extremely important. It implies that the critical fluctuations do
not destroy the second--order phase transitions, at least, if the
anisotropy of the initial Hamiltonian is not too strong. Since the
stable fixed point is located on the plane $v=z$ it is certainly
relevant to the critical behavior of $Tb Au_2$, $Dy C_2$, $Tb D_2$,
and $Nd$.

Let us note, that the $\ve$--expansions of the eigenvalue exponents
$\la_1$ and $\la_2$ for the II--tetragonal point substantially differ
from the others. Namely, their series prove to be shorter by one
order (see Table 1 and Table 2). This phenomenon originates from
the two--fold degeneracy of the roots of the characteristic
polynomial in the one--loop approximation. As a consequence,
the eigenvalue exponents should be expanded in $\sqrt{\ve}$ rather
than $\ve$. It can be shown, however, that for almost all $N$
non--integer powers drop from $\la_1$ and $\la_2$ for the eighth
fixed point in every order of the perturbation theory. For the
special case $N=2$, significant from the physical viewpoint, one
cannot make such a statement within three--loop approximation. To
answer that question one should take into account at least
four--loop contributions. Apart from whether or not non--integer
powers of $\ve$ appear in the expansions, one--loop degeneracy of
$\la$'s results in reduction of information available from a given
approximation. So, assuming that $\sqrt{\ve}$ will not appear in
the series for $N=2$, evaluating coefficients of $\la$'s in third
order in $\ve$
would require accounting
five--loop terms. To understand
the structure of the eigenvalue exponent series with one--loop
degeneracy, we conduct detailed analysis of the problem in
Appendix B.

\section{Marginal dimensionality and critical exponents}
\label{sec:4}

We have shown in the previous section that for the physically
interesting cases $N=2$ and $N=3$ the II--tetragonal fixed point
is three--dimensionally stable in $3D$. The question may be put
forward whether this point is stable for all $N$. To answer it
the critical dimensionality of the order parameter $N_c$ needs
to be calculated.  It separates two different regimes
of critical behavior of the model. When $N>N_c$ the II--tetragonal
rather than the Bose fixed point is three--dimensionally stable in
$3D$. At $N=N_c$ they interchange their stability so that when
$N<N_c$ the stable fixed point is the Bose one. The
$\ve$--expansion for $N_c$ can be found from the condition
$v_c=z_c=0$ imposed on the coordinates of the eighth fixed
point (see Sec. \ref{sec:3}). Three--loop approximation gives
\be
 N_c&=& 2 - 2 \ve + {5\over 6}(6 \zt(3)-1) \ve^2+ O(\ve^3),
 \label{eq:4.1}
\ee
where $\ve={1\over 2}$ corresponds to the physical space
dimensionality $D=3$. The same expansion holds within the
plane $(u,v)$. Note, that expression (\ref{eq:4.1}) coincides
with that found for the cubic model with the complex order
parameter \cite{7}. Such a coincidence is not occasional because,
as was already emphasized, for $v=z$ the model (\ref{eq:1.1})
goes over into to the complex cubic model.

Unfortunately, RG expansion (\ref{eq:4.1}) is known to be
divergent. Nevertheless, the physical information may be extracted
from it, provided some resummation  method is applied. Since the
series of $N_c$ is alternating, the Borel transformation combined
with its proper analytical continuation may play a role of such
method. To perform analytical continuation the Pad$\acute {\rm e}$
approximant only of the type  $[1/1]$ may be used within given
approximation. The Pad$\acute {\rm e}$--Borel summation of the
expansion (\ref{eq:4.1}) gives
\be
 N_c&=& a - {2b^2\over c} + {4b^3\over {c^2\ve}}
 \exp\Biggl(- {2b\over {c\ve}}\Biggr)
 E_i\Biggl({2b\over {c\ve}}\Biggr), \label{eq:4.2}
\ee
where $a$, $b$, $c$ are the coefficients before $\ve^0$, $\ve^1$,
$\ve^2$ in Eq.(\ref{eq:4.1}), respectively, and $E_i(x)$ is the
exponential integral. Setting $\ve={1\over 2}$ in
Eq.(\ref{eq:4.2}), we obtain the value of the critical
dimensionality
\be
 N_c&=& 1.50 \>.
 \label{eq:4.3}
\ee
This number is close to $N_c=1.47$ found within the three dimensional
RG approach \cite{6}. Since $N_c$ lies below two, the critical
behavior of antiferromagnets ($N=2$, $N=3$) and $Nb O_2$ ($N=2$)
must be governed by the II--tetragonal fixed point.

Now let us turn to calculating the critical exponents. To this end,
substitute the coordinates of fixed points (see Sec. \ref{sec:3})
into the expressions for $\gm^{-1}$ and $\eta$ (Eqs. (\ref{eq:2.7})
and (\ref{eq:2.8})). For the stable fixed point 8 it gives
\be
\gm^{-1}&=&1+\frac{\ve}{(5N-4)}3(1-N) + \frac{\ve^2}{(5N-4)^3}
(N-1)(40N^2 - 214 N + 205)
\nn\\&+&
\frac{\ve^3}{(5N-4)^5}(1-N)(12 \zt(3)(5N-4)(32 N^3 - 156 N^2
+ 159 N - 13)
\nn\\&-&
 940 N^4 - 6748 N^3 +42681 N^2 -67102 N + 32558),
\label{eq:4.4}\\
\eta&=&\frac{\ve^2}{(5N-4)^2}(N-1)(2N-1)
+ \frac{\ve^3}{2(5N-4)^4}(N-1)(190N^3-535 N^2
\nn\\
&+&652 N - 324)+
\frac{\ve^4}{4(5N-4)^6})(1-N)(96 \zt(3) (5N-4)(32N^4-128 N^3
\nn\\
&+&
212 N^2-153 N + 33)-10570 N^5 +22691 N^4 +68527 N^3
\nn\\
&-&280399 N^2+326888 N - 127676).
\nn
\ee
From Eqs.(\ref{eq:4.4}) we find for $N=2$
\be
\gm^{-1}&=&1-\frac{\ve}{2} - \frac{7\ve^2}{24}+
\frac{\ve^3(84\zt(3)-1)}{144},
\nn\\
\eta&=&\frac{\ve^2}{12} + \frac{5\ve^3}{36}
+\frac{\ve^4(13-21\zt(3))}{108}
\label{eq:4.5}\ee
and for $N=3$
\be
\gm^{-1}&=&1-\frac{6\ve}{11} - \frac{14\ve^2}{121}
+\frac{2\ve^3(912\zt(3)+3905)}{14641},
\nn\\
\eta&=&\frac{10\ve^2}{121} + \frac{177\ve^3}{1331}
+\frac{\ve^4(50083-59328\zt(3))}{322102},
\label{eq:4.6}\ee
where $\ve={1\over 2}$ as before.
Other critical exponents are found from the well known
scaling relations.

We will focus first on qualitative discussion of the results
obtained. As was found in Ref. \cite{2}, the critical
exponents of the Heisenberg and the II--tetragonal fixed points
coincide within the two--loop approximation. Three--loop
analysis yields for the Heisenberg fixed point at $N=2$
\be
\gm^{-1}&=&1-\frac{\ve}{2} - \frac{7\ve^2}{24}
+\frac{\ve^3 (28\zt(3)-11)}{48},
\nn\\
\eta&=&\frac{\ve^2}{12} + \frac{5\ve^3}{36}
+\frac{\ve^4(13-21\zt(3))}{108} .
\label{eq:4.7}\ee
Comparing (\ref{eq:4.7}) with (\ref{eq:4.5}) we see that the critical
exponent $\gamma$ of the Heisenberg and the II--tetragonal fixed
points is different in third order in $\ve$, although that difference
is not too strong. This is one  of the results of our investigation.

It is known that RG series for critical exponents are badly
divergent. However they contain important  physical information which
can be extracted provided some procedure making them convergent is
applied. Unfortunately, it is impossible to use simple
Pad$\acute{\rm e}$--Borel summation to process series
(\ref{eq:4.5})--(\ref{eq:4.7}) because their coefficients have
irregular signs, in contrast to  the critical dimensionality $N_c$.
The most appropriate resummation scheme known for now is a
modification of the Borel technique. Principal underlying ideas of
this method are the analytical continuation of the Borel transform
beyond its circle of convergence over the  cut--plane and a conformal
mapping sending the cut--plane onto the circle. Such an operation
leads to integration of a holomorphic function represented by an
absolutely convergent series and allows to perform integration prior
to summation, thus  substantiating the perturbation theory approach.
The algorithm just mentioned incorporates both exactly calculated
first several terms  and high order asymptotic behavior of
perturbation series. For the simple $O(n)$--symmetric model the
coefficients at large order $k$ were shown to look like
$(-1)^k k! a^k k^b $ \cite{29,30}. It can be expected that in
complex models  with more than one coupling constants asymptotics of
RG series will comprise such a factor, at least. Parameters $a$ and
$b$ play an essential role in the modified Borel method.
For a given series
$$ F(\ve)=\sum f_k \ve^k$$
transformation
$$ F(\ve)\sim \int_0^\infty e^{-{t\over{a \ve}}}
\Biggl({t\over{a \ve}}\Biggr)^b
 B(t) {dt\over{a \ve}},$$
 where $B(t)=\sum_k\frac{f_k}{a^k \Gamma(b+k+1)} t^k$,
 is followed by the conformal mapping
 $$\omega=\frac{\sqrt{t+1}-1}{\sqrt{t+1}+1}.$$
Function $B(t)$ is represented by the series in $\omega$
$$B(t(\omega))=\Biggl(\frac{2}{1-\omega}\Biggr)^{2\la}\sum_{k}
 A_k(\la) \omega^{k}$$
where the additional parameter $\la$ is introduced to eliminate
possible singularity at $\omega=1$. Since the type of that
singularity is unknown $\la$ is chosen so as to ensure the most
rapid convergence of the series \cite{26}.

The main obstruction for application  of the method just outlined
to the model (\ref{eq:1.1}) as well as to a great deal of other
complex anisotropic systems is unknown asymptotic parameters $a$
and $b$. Evaluating them requires enormous efforts. In the case of
the $n$--vector model with one coupling constant the  parameters
$a$ and $b$ have been exactly calculated that allowed to obtain
accurate numerical estimates for the critical exponents
\cite{22,26,31,32}. Attempts to find asymptotic parameters for the
cubic model also were made \cite{33}. They proved to be successful,
however, only within the assumption of very weak anisotropy. Despite
there is no information about asymptotic parameters of the model
(\ref{eq:1.1}) available at the moment we chose to resort to the
resummation scheme of \cite{22,26}, in view of the following
arguments. Although the asymptotic parameters for the isotropic
model are explicitly calculated, in Ref.\cite{22} parameter $b$ was
varied in a neighborhood of the exact value. It is justified by that
exact $a$ and $b$ determine large order behavior of $F(\ve)$ while
actually one deals with only few terms of  perturbation series. We
believe therefore that, in connection with the model (\ref{eq:1.1}),
similar manipulations may be valid not only with respect to parameter
$b$ but to parameter $a$ as well. Variation of $a$ and $b$ in a range
containing exact asymptotic parameter values of the $O(n)$--symmetric
fixed point and using $\la$ as an optimizing parameter result in
values of the critical exponents displayed in the Table \ref{Tab3}.
Here we suppose that unknown exact asymptotic values $a$ and $b$ of
the model  (\ref{eq:1.1}) are not much distant from those of the
$O(n)$--symmetric model. The error of the numerical estimates
is established  through the dispersion of the output due to the
variation of $a$, $b$, and $\la$.
\begin{table}
\caption{Critical exponents $\eta$ and $\gm$ of the model (1.1) for
$N=2$ and $N=3$ calculated within the three--loop approximation}
\label{Tab3}
\vspace{0.5cm}
\hspace{0.5cm}
\begin{tabular}{|c|l|l|l|l|}\hline
Type of             &\multicolumn{2}{|c|}{$N=2$}&
                     \multicolumn{2}{|c|}{$N=3$}
                   \\[0pt]     \cline{2-5}
        fixed point &\multicolumn{1}{|c|}{$\eta$}
                    &\multicolumn{1}{|c|}{$\gamma$}
                    &\multicolumn{1}{|c|}{$\eta$}
                    &\multicolumn{1}{|c|}{$\gamma$}
                   \\[0pt]  \hline
Heisenberg   &$0.0285\pm0.0002 $&$1.368\pm0.004$
                    &$0.0271\pm0.0002 $&$1.440\pm0.005$
                   \\[0pt] \hline
Bose         &$0.0279\pm0.0002 $&$1.265\pm0.011$
                    &$0.0279\pm0.0002 $&$1.265\pm0.011$
                   \\[0pt] \hline
II-tetragonal&$0.0285\pm0.0002 $&$1.355\pm0.015$
                    &$0.0281\pm0.0002 $&$1.380\pm0.008$
                   \\[0pt]\hline
\end{tabular}
\end{table}

As may be seen from the  table, the critical exponents of the
II--tetragonal fixed point appear to be close to those of the
Heisenberg point. Unfortunately, we cannot compare the critical
exponent values of the II--tetragonal fixed point with their
two--loop analogs and therefore decide how far they shift from the
Heisenberg ones with higher--loop terms being taken into account.
The point is that the estimates of the critical exponents were
done in Ref. \cite{2} by direct summation of the $\ve$--expansion
terms setting
$\ve=1$, that was illegal for the  asymptotic series. Under such
circumstances, let us compare the results obtained with predictions
given for the investigated model by the RG procedure in three
dimensions. Two-- and three--loop calculations carried out in
Refs. \cite{12,14} shown that the critical exponents of the
II--tetragonal point turned out to be close not to those of the
$O(2N)$--symmetric model, as in the case of the $\ve$--expansion
method, but to the exponents of the $3D$ $XY$ (Bose) model. Within
the RG analysis in $3D$ it is a consequence of the closeness of the
stable fixed point 8 and the Bose point 5 on the three dimensional
RG flow diagram. Despite of such a distinction in
estimates of the critical exponents given by these two RG approaches,
one can hope that involving higher perturbation orders and using an
appropriate resummation technique will soften this discrepancy.
Not so strong difference between the critical exponent values
for the II--tetragonal fixed point obtained within the
$\ve$--expansion and the $3D$  RG  methods  may serve as a
possible confirmation to this conjecture. Indeed, $3D$ RG  analysis
of the model (\ref{eq:1.1}) yields the following estimates for the
critical exponents of the II--tetragonal fixed point:
$\gm=1.336$, $\eta= 0.0261$ at $ N=2 $ and
$\gm=1.329$, $\eta= 0.0261$ at $ N=3 $. Comparing these numbers with
their analogs from Table  \ref{Tab3}, we conclude that the relative
deviation does not exceed  $4\%$ for $\gm$ and $8\%$ for $\eta$, that
is not so bad for the three--loop approximation. An additional
stimulus for our hope is the beautiful agreement of numerical
estimates of the critical exponents for the simple $O(n)$--symmetric
model achieved in sufficiently high orders of the perturbation theory
between the $3D$ RG \cite{22,31} and $\ve$--expansion \cite{32}
approaches. So, for the Ising model the magnetic
susceptibility exponent was found to be $\gm=1.241$ in the frame of
$3D$ RG and $\gm=1.239$ within the $\ve$--expansion  method. The
relative deviation of these values is about $0.1\%$.

At last, we would like to emphasize, that although the accuracy of
the estimates of the critical exponents achieved in the paper cannot
be regarded as satisfactory the numerical values of the critical
exponents for the II--tetragonal fixed point presented here are, in
our opinion, the most realistic among those so far obtained on the
base of the $\ve$--expansion method.
\section{Conclusion}

The complete RG analysis of a model with three quartic coupling
constants and $2N$--component real order parameter field describing
phase transitions in certain cubic and tetragonal antiferromagnets as
well as  the structural phase transition in $NbO_2$ crystal  has been
carried out within the three--loop approximation in
$D=4-2\ve$
dimensions. Perturbation expansions for the $\bt$--functions of the
record length were obtained using dimensional regularization and the
minimal subtraction scheme. Coordinates of the fixed points and their
eigenvalue exponents were calculated for arbitrary $N$. The analysis
performed for the eigenvalue exponents has shown that for $N\geq 2$
the II--tetragonal rather than the Bose fixed point is absolutely
stable in the physical space within given approximation.
The three--loop $\ve$--expansion
for the critical dimensionality of the order parameter $N_c$ was
found and processed by the Pad{$\acute{\rm e}$}--Borel resummation
technique. The numerical estimate $N_c=1.50$ obtained confirms the
conclusion about the stability of the II--tetragonal fixed point.
Consequently, the phase transitions in the $NbO_2$ crystal and
antiferromagnets $TbAu_2$, $DyC_2$, $K_2 Ir Cl_6$, $Tb D_2$, and
$Nd$  are of second order and their critical thermodynamics should
be controlled by this point, in the frame of given approximation.

It was observed that the degeneracy of the eigenvalue exponents in
the one--loop approximation for the II--tetragonal fixed point
resulted in certain difficulties in calculating their $\ve$--series.
According to the analysis carried out, two--fold degenerate
eigenvalue exponents should be expanded not in $\ve$ but in
$\sqrt{\ve}$. Although non--integer powers of $\ve$ was shown to
drop from the expansions for all $N$ excepting $N=1,2$, such a
degeneracy led to reduction in length of the RG series for
eigenvalue exponents and therefore to the loss of accuracy
expected from given approximation. Indeed, within the three--loop
approximation we actually obtain two--loop--like pieces of the
series and evaluation of the term of order $\ve^3$
may require to
account the five--loop contributions. To understand the structure
of the eigenvalue exponent series for the special case $N=2$,
important physically, one has to consider at least four--loop
approximation.

Perturbation expansions for the critical exponents $\gm$ and $\eta$
were calculated up to $\ve^3$ and $\ve^4$, respectively. For $N=2$
the magnetic susceptibility exponents for the II--tetragonal and
Heisenberg fixed points were found to be different in third order
in $\ve$. For the first time  within the $\ve$--expansion method
the numerical estimates of the critical exponents of the model under
consideration were given on the base of the Borel summation technique
modified with a conformal mapping. For the physically interesting
cases $N=2$ and $N=3$ the critical exponents of the II--tetragonal
fixed point turned out to be numerically close to those of
the Heisenberg one.
On the contrary, in the frame of the field--theoretical RG approach
in three dimensions the critical exponents of the Bose and the
unique stable fixed points are close to each other.
Possibly, these two alternative RG approaches will be in better
agreement, provided the higher--loop contributions are taken into
account.

The results achieved in our study seem to be self--consistent
although there is definite  discrepancy with
the non--perturbative theoretical predictions. We believe that
it is the effect of insufficiently high approximation employed
and the problem of bringing the results given by the
$\ve$--expansion method into accordance with those of other
theoretical approaches and experimental data needs to be solved.

\vspace{0.5cm}
{\Large\bf  Acknowledgement}
\vspace{0.5cm}

We are grateful to Prof.~J.~Zinn--Justin for interesting remarks
concerning the property of the stable fixed point made at
the International conference {\em Renormalization Group '96}.
We would like to thank Prof.~A.~I.~Sokolov and
Dr.~B.~N.~Shalaev for numerous illuminating
discussions. One of us (K.~B.~V.) have benefitted from
conversations with Dr.~I.~O.~Mayer and Prof.~A.L.Korzhenevskii.
This work was supported in part by Russian Research Program
"Fullerenes and Atomic Clusters" (grant 94024).
\appendix
\section{Appendix}
\label{A}
In this appendix we present the eigenvalue exponents of all the
fixed points for arbitrary $N$.\\[6pt]
\begin{tabular}{llll}
{\bf 1.}&\multicolumn{3}{l}{\bf  Gaussian fixed point}
\\[6pt]
&$\la_u$&=&$\la_v=\la_z=\ve$.
\\[6pt]
\end{tabular}
\\
\begin{tabular}{llll}
{\bf 2.}&\multicolumn{3}{l}{\bf $O(2N)$--symmetric or Heisenberg
fixed point}
\\[6pt]
&$\la_u$&=&$-\ve+\frac{3}{(N+4)^2}(3N+7) \ve^2 -
\frac{1}{2(N+4)^4}\Bigl(48 \zt(3) (N+4)(5N+11)+
$\\[6pt]&&&$
33N^3+269N^2+1072N+1196\Bigr) \ve^3$,
\\[6pt]
&$\la_v$&=&$\la_z=\frac{1}{N+4}(N-2) \ve+\frac{1}{(N+4)^3}
(5N^2+7N+38) \ve^2-
$\\[6pt]&&&$
\frac{1}{2(N+4)^5}\Bigl(48 \zt(3)(N+4)(2N^2+7N+23)-
13N^4-199N^3-
$\\[6pt]&&&$
2(183N^2+98N-532)\Bigr)\ve^3$ .
\\[6pt]
\end{tabular}
\\
\begin{tabular}{llll}
{\bf 3.}&\multicolumn{3}{l}{\bf Ising fixed point}
\\[6pt]
&$\la_u$&=&$\la_z={1 \over 3} \ve-{38 \over 81} \ve^2
+{1\over 2187}(2592 \zt(3)-937)\ve^3$ ,
\\[6pt]
&$\la_v$&=&$-\ve+{34 \over 27}\ve^2
-{1 \over 729}(2592 \zt(3)+1603) \ve^3$ .
\\[6pt]
\end{tabular}
\\
\begin{tabular}{llll}
{\bf 4.}&\multicolumn{3}{l}{\bf Cubic fixed point}\\[6pt]
&$\la_1$&=&$-\ve+\frac{2N-1}{27N^2(N+1)}(17N^2-2N+53)\ve^2-
$\\[6pt]&&&$
\frac{1}{1458N^4(N+1)^3}
\Bigl(1296 \zt(3)N(4N^6+4N^4+27N^3+15N^2-11N-7)+
$\\[6pt]&&&$
3206N^7-11683N^6+48012N^5+34522N^4-111830N^3+71205N^2+
$\\[6pt]&&&$
3452N-11236\Bigr)\ve^3$ ,
\\[6pt]
&$\la_2$&=&$\frac{2-N}{3 N}\ve+\frac{2N-1}{81N^3(N+1)}
(19N^3-36N^2-165N+106)\ve^2+
$\\[6pt]&&&$
\frac{1-2N}{4374N^5(N+1)^3}
\Bigl(1296 \zt(3) N(2N^6+3N^5-22N^4-39N^3-3N^2+20N+
$\\[6pt]&&&$
7)-937N^7+7850N^6-
40674N^5+6832N^4+146287N^3-99642N^2-
$\\[6pt]&&&$
27196N+22472\Bigr)\ve^3$ ,
\\[6pt]
&$\la_z$&=&$\frac{N-2}{3N}\ve+\frac{1-2N}{81N^3}
(19N^2-127N+106)\ve^2+
$\\[6pt]&&&$
\frac{1}{4374 N^5}\Bigl(1296 \zt(3)N(4N^4-20N^3+4N^2+21N-7)
-1874N^5-
$\\[6pt]&&&$
9997N^4+94159N^3-168626N^2+109028N-22472\Bigr)\ve^3$ .
\\[6pt]
\end{tabular}
\\
\begin{tabular}{llll}
{\bf 5.}&\multicolumn{3}{l}{\bf Bose fixed point}
\\[6pt]
&$\la_u$&=&${1 \over 5}\ve-{14 \over 25}\ve^2
+{1 \over 625}(768\zt(3)-311)\ve^3$ ,
\\[6pt]
&$\la_1$&=&$-{1 \over 5}\ve+{2 \over 5} \ve^2
-{1 \over 625}(768 \zt(3)+29)\ve^3$ ,
\\[6pt]
&$\la_2$&=&$-\ve+{6 \over 5}\ve^2
-{1 \over 125}(384 \zt(3)+257)\ve^3$ .
\\[6pt]
\end{tabular}
\\
\begin{tabular}{llll}
{\bf 6.}&\multicolumn{3}{l}{\bf VZ--cubic fixed point}
\\[6pt]
&$\la_u$&=&$\la_1={1 \over 3}\ve-{38 \over 81}\ve^2
+{1 \over 2187}(2592 \zt(3)-937)\ve^3$ ,
\\[6pt]
&$\la_2$&=&$-\ve+{34 \over 27}\ve^2
-{1 \over 729}(2592 \zt(3)+1603)\ve^3$ .
\\[6pt]
\end{tabular}
\\
\begin{tabular}{llll}
{\bf 7.}&\multicolumn{3}{l}{\bf I-tetragonal fixed point}
\\[6pt]
&$\la_1$&=&$\frac{N-2}{3 N}\ve+\frac{1-2 N}{81 N^3}
(19 N^2-127 N+106) \ve^2+
$\\[6pt]&&&$
\frac{1}{4374 N^5}\Bigl(1296 \zt(3) N
(4 N^4-20 N^3+4 N^2+21 N-7)-
$\\[6pt]&&&$
1874 N^5-9997 N^4+94159 N^3-168626 N^2
+109028 N-22472\Bigr) \ve^3$ ,
\\[6pt]
&$\la_2$&=&$\frac{2-N}{3 N} \ve+\frac{2 N-1}{81 N^3 (N+1)}
(19 N^3-36 N^2-165 N+106) \ve^2+
$\\[6pt]&&&$
\frac{1-2 N}{4374 N^5 (N+1)^3} \Bigl(1296 \zt(3) N
(2 N^6+3 N^5-22 N^4-39 N^3-3 N^2+20 N+
$\\[6pt]&&&$
7)-937 N^7+7850 N^6-40674 N^5+6832 N^4+146287 N^3-99642 N^2-
$\\[6pt]&&&$
27196 N+22472\Bigr) \ve^3$ ,
\\[6pt]
&$\la_3$&=&$-\ve+\frac{2N-1}{27N^2(N+1)}(17N^2-2N+53) \ve^2-
$\\[6pt]&&&$
\frac{1}{1458 N^4 (N+1)^3}
\Bigl(1296 \zt(3) N (4 N^6+4 N^4+27 N^3+15 N^2-11 N-7)+
$\\[6pt]&&&$
3206 N^7-11683 N^6+48012 N^5+34522 N^4-
111830 N^3+71205 N^2+
$\\[6pt]&&&$
3452 N-11236\Bigr) \ve^3$   .
\\[6pt]
\end{tabular}
\\
\begin{tabular}{llll}
{\bf 8.}&\multicolumn{3}{l}{\bf II-tetragonal fixed point}
\\[6pt]
&$\la_1$&=&$\frac{2-N}{5 N-4}\ve+\frac{1-N}{(5 N-4)^3 (2 N-1)}
\Bigl(4 sign(N-1) |5 N^3+6 N^2-48 N+32|-
$\\[6pt]&&&$
3 (40 N^3-208 N^2+253 N-66)\Bigr) \ve^2$ ,
\\[6pt]
&$\la_2$&=&$\frac{2-N}{5 N-4}\ve+\frac{(N-1)}{(5 N-4)^3 (2 N-1)}
\Bigl(4 sign(N-1) |5 N^3+6 N^2-48 N+32|+
$\\[6pt]&&&$
3 (40 N^3-208 N^2+253 N-66)\Bigr)\ve^2$  ,
\\[6pt]
&$\la_3$&=&$
-\ve+
\frac{1}{(5 N-4)^2 (2 N-1)}(60 N^3-160 N^2
+181 N-85) \ve^2+
$\\[6pt]&&&$
\frac{1}{2 (5 N-4)^4 (1-2 N)^3}\Bigl(48 \zt(3) (2 N-1)^2
(5 N-4)(32 N^4-128 N^3+212 N^2-
$\\[6pt]&&&$
153 N+33)+20560 N^7-165328 N^6+644392 N^5-1406864 N^4+
$\\[6pt]&&&$
1756745 N^3-1224341 N^2+433704 N-59052\Bigr)
\ve^3$ .
\end{tabular}\\
Here $\ve={1\over 2}$ corresponds to the physycal space.

\section{Appendix}
\label{B}
As was noted in Sec. \ref{sec:3}, the II--tetragonal fixed point
has an unusual structure of the series of the $\ve$--expansion for
the eigenvalues of the stability matrix. Namely, those series are
shorter by one order, comparing to their analogs for the other
fixed points. This is, actually, a consequence
of the multiplicity of the roots of the characteristic equation in
the one--loop approximation, that may cause non--integer powers of
$\ve$ to contribute to the expansions. Such a conclusion seems so
exotic that deserves thorough investigation, to which the present
Appendix is devoted. The result of the analysis is that for
$N\not=1,2$ non--integer powers do not appear in the eigenvalue
exponents series in all orders of the perturbation theory. As to the
physically important case $N=2$, we cannot make such a statement
within the  three--loop approximation. To answer the question
whether or not non--integer powers of $\ve$ will appear in the
expansions the higher--loop (at least four--loop) approximations
need to be considered.

An eigenvalue $\la$ of the stability matrix is a root of its
characteristic polynomial. It is convenient, rather, to deal with
the quantity $y={\la\over \ve}$ which is a root of the corresponding
reduced polynomial denoted hereafter $P(y,\ve)$. In every order of
the perturbation theory its coefficients are also polynomials in
$\ve$, therefore a piece of the series of $y(\ve)$ determined
within corresponding approximation coincides with that of some
algebraic function. Such a function is not analytical in those
points on the complex plane where the defining polynomial
($P(y,\ve)$ in our case) has multiple roots. Instead, it has
branching of an order not greater than the multiplicity of the
root. As to the II--tetragonal point, at $\ve=0$ (one--loop
approximation) the reduced characteristic polynomial has two
equal roots of the three. It leads to the conclusion that
$y(\ve)$ should be expanded not in $\ve$ but in $\sqrt\ve$ as a
Puiseux series \cite{35}. That is how half--integer powers of
$\ve$ may occur in the series of eigenvalue exponents. Let us
show, however, that in the model
(\ref{eq:1.1})
they are absent at
least for $N\not=1,2$. Consider the reduced characteristic equation
\be
 - y^3 + a y^2 -  b y + c &=& 0 \label{eq:B.1}
\ee
and assume
\be
a&=&a_0+a_1 \ve+a_2 \ve^2+\ldots ,\nn\\
b&=&b_0+b_1 \ve+b_2 \ve^2+\ldots , \label{eq:B.2}\\
c&=&c_0+c_1 \ve+c_2 \ve^2+\ldots ,\nn\\
y&=&y_0+y_{1\over 2} \ve^{1 \over 2}+y_1 \ve+y_{3\over 2}
\ve^{3 \over 2}+\ldots .\nn
\ee
Here we have omitted higher terms relevant to higher than three
loops. The coefficients in (\ref{eq:B.2}) are rational functions
in $N$:
\begin{equation}
\begin{array}{lll}
a_0&=&\frac{-1}{(5N-4)}(7 N - 8)     ,\nn\\
a_1&=&\frac{1}{(5N-4)^3}(270 N^3 - 1129 N^2 + 1591 N - 736 ) ,\nn\\
a_2&=&\frac{-1}{2 (5N-4)^5}(48 \zt(3)(5 N - 4)(144 N^4 - 720 N^3
+ 1289 N^2 - 947 N +230) \nn\\
   &+& 10030 N^5 - 104229 N^4 + 429747 N^3 - 804632 N^2
   + 691620 N - 222720) ,\nn\\
b_0&=&\frac{1}{(5N - 4)^2}(N - 2)(11 N - 10)    ,\nn\\
b_1&=&\frac{-2}{(5N - 4)^4}  (510 N^4 - 3157 N^3+6615 N^2
- 5832 N + 1868)     ,\nn\\
b_2&=&\frac{1}{(5N - 4)^6}(48 \zt(3)(5 N - 4) (272 N^5 - 1824 N^4
+ 4455 N^3 - 5095 N^2 \nn\\
   &+& 2754 N -558) +  25890 N^6 - 338437 N^5 + 1547050 N^4
   - 3437182 N^3 \nn\\
   &+& 4044203 N^2  - 2430752 N + 589412)     ,\nn\\
c_0&=&\frac{-1}{(5N - 4)^2}(N - 2)^2     ,\nn\\
c_1&=&\frac{1}{(5N - 4)^4}(N - 2)(150 N^3 - 809 N^2 + 1229 N
- 566)     ,\nn\\
c_2&=&\frac{-1}{2 (5N - 4)^6}(48 \zt(3)(N - 2)(5 N - 4)(80 N^4
- 464 N^3 + 865 N^2 \nn\\
   &-&  641 N + 164)  + 13950 N^6 - 184745 N^5 + 887705 N^4
   - 2072060  N^3 \nn\\
   &+&     2541094  N^2 - 1575640 N +389512)           .\nn
\end{array}\label{eq:B.3}
\end{equation}

Substituting (\ref{eq:B.2})  into  (\ref{eq:B.1}) and letting
$\ve=0$ we find that $y_0$ is two--fold degenerate taking
the value $-1$ once
and $\frac{2-N}{5N-4}$ twice.
Comparing factors before equal powers of $\ve$ in
(\ref{eq:1.1}), we
recursively evaluate next terms. The first appearance of a
coefficient $y_l$, $l > 0$, occurs at the $l$-th step, where it has
the multiplier $\partial_y P(y_0,0)$. Factor $\partial_y P(y_0,0)$
vanishes due to the multiplicity of $y_0$, hence $y_l$ actually
cannot be determined from the $l$-th order. So, for $y_{1\over 2}$
from first order in $\ve$ we have the quadratic equation
\be
y^2_{1\over 2}&=&\frac{a_1 y_0^2 - b_1 y_0+c_1}{3 y_0 - a_0}.
\label{eq:B.4}
\ee
The denominator on the right--hand side is non--zero because it is
proportional to $\partial^2_y P(y_0,0)$ and only two of the three
roots coincide.
Substitution  of (\ref{eq:B.3}) into  (\ref{eq:B.4}) gives
$y_{1\over 2}=0$. The next order ($\ve^{3\over 2}$) does not provide
$y_1$, as it might be expected, because at this step  equation
(\ref{eq:B.1}) vanishes identically. Considering factors before
$\ve^2$, we come to the quadratic equation for $y_1$:
\be
  y^2_1 (-3 y_0+a_0) + y_1 (2 a_1 y_0-b_1) + a_2 y^2_0
  - b_2 y_0 + c_2=0. \label{eq:B.5}
\ee
The highest coefficient $-3 y_0+a_0$ is proportional to the
non--zero quantity $\partial^2_y P(y_0,0)$, therefore $y_1$ is
explicitly determined:
\be
 y_1&=&\frac{3(N-1)(40N^3-208N^2+253N-66)}{(2N-1)(5N-4)^3} \nn\\
    &\pm& \frac{4 |(N-1)(N-2)(N+4)(5N-4)|}{(2N-1)(5N-4)^3}.
     \label{eq:B.7}
\ee
Factors before $\ve^{5\over 2}$ in Eq. (\ref{eq:B.1}) obey the
equation $$B y_{3\over 2} =0,$$ where
\be
  B=-6 y_0 y_1+2 a_0 y_1+ 2 a_1 y_0 - b_1. \label{eq:B.6}
\ee
We have come to the crucial point of the consideration. Supposing
$B\not=0$ we have $y_{3\over 2} =0$. Further, this step can be put
into the base of mathematical induction in proving  disappearance
of non--integer powers of $\ve$. For $m\geq {3\over 2}$  we have
$$ 0 = \partial_y P(y_0,0) y_{m+1} +
(- 6 y_0 + 2 a_0) y_{1\over 2} y_{m+{1\over 2}}  +
B y_{m}+... ,$$
were the terms depending only on $y_l$ with $l<m$ have been
suppressed. The first two terms turn into zero while $B\not =0$,
so the coefficients  $y_m$, $m\ge {3\over 2}$, can be calculated
recurrently. Suppose now that $m={{2k+1} \over 2}$ with $k$
integer. The equation on $y_{{2k+1} \over 2}$ can be written in
the form $$B y_{{2k+1} \over 2}=F(y_{1 \over 2},y_{3\over 2},
\ldots, y_{{2k-1} \over 2}),$$ where $F$ is some polynomial with
zero absolute term in it (because $a(\ve)$, $b(\ve)$, and $c(\ve)$
does not depend upon non--integer powers). Recursively we have
$y_{{2k+1} \over 2}=0$ as required.

Thus, the multiplicity of the roots does not give rise to
non--integer powers of $\ve$ unless expression (\ref{eq:B.6})
vanishes. The important fact is that (\ref{eq:B.6}) is the
derivative of the quadratic polynomial (\ref{eq:B.5}) with respect
to $y_1$. It implies that (\ref{eq:B.6}) turns into zero if and
only if $y_1$ is two--fold degenerate. Formula (\ref{eq:B.7})
shows that this possibility is realized only for $N=1$ and $N=2$.
Concerning these two special cases, one can see, that
$y_{3\over 2}$ is determined from  a quadratic equation
$\partial^2_y P(y_0,0) (y_{3\over 2})^2 =\ldots$ arising from
comparing factors before $\ve^3$ in Eq. (\ref{eq:B.1}).
The right--hand side depends on the four--loop contributions.
So, if it does not vanish, the expansions would contain non--integer
powers. Otherwise, $y_{3\over 2}=0$ and the five--loop approximation
gives, in its turn, a quadratic equation for $y_2$.

In summary, let us have a look at the structure of the one--loop
degenerate eigenvalue exponent series in general. For every $l$
let $d(l)$ be the order of the expansion of the reduced
characteristic equation in which $y_l$ is determined. It ranges
from $l+1$ to $2l$. If  $d(l)=2l$ then $y_l$ is found from a
quadratic equation with the non--zero highest coefficient
$\partial^2_y P(y_0,0)$. We shall say that the solution
$y(\ve)$ {\em splits at the step} $l_s$ if that equation gives
two different values of $y_{l_s}$. If $y(\ve)$ does not split at
all, it is convenient to assign $l_s\equiv \infty$. Let us
formulate the resulting theorem as the set of four propositions.
\\[12pt]
{\bf Theorem}
{\em
\begin{enumerate}
\item
Either the characterictic  equation has
two equal roots in every order of the perturbation theory or
its  solution splits at a finite step $l_s$.
For every half--integer number $l$ from the interval $[0,l_s]$
coefficient $y_l$ is determined in the order $d(l)=2l$.
\item
In the case of finite $l_s$ coefficient $y_{l_s+m}$ is determined
in the order $d(l_s+m)=2l_s+m$ for all $m\geq {1\over 2}$.
\item
Coefficients $y_l$ with non--integer numbers $l < l_s$ are equal
to zero.
\item
Non--integer powers of $\ve$ contribute to the expansions
of the eigenvalue exponents if  and only if $l_s$ is a non--integer
number.
\end{enumerate}
}
We have demonstrated how the theorem works in the frame of the model
under consideration, and its full proof will be given elsewhere.


\begin{thebibliography}{37}
\bibitem{1}D. Mukamel, Phys. Rev. Lett. {\bf 34}, 481 (1975).
\bibitem{1a}D. Mukamel and S. Krinsky, J. Phys. C {\bf 8},
L496 (1975).
\bibitem{2}D. Mukamel and S. Krinsky, Phys. Rev. B {\bf 13},
5078 (1976).
\bibitem{3}P. Bak and D. Mukamel, Phys. Rev. B {\bf 13},
5086 (1976).
\bibitem{4}S. A. Brazovskii, I. E. Dzyaloshinskii,
and B. G. Kukharenko, Zh. Eksp. Teor. Fiz. {\bf 70},
2257 (1976)
[Sov. Phys. JETP {\bf 43}, 1178 (1976)].
\bibitem{5}E. J. Blagoeva et al., Phys. Rev. B {\bf 42},
6124 (1990).
\bibitem{6}S. A. Antonenko and A. I. Sokolov, Phys. Rev. B
{\bf 49}, 15901 (1994).
\bibitem{7}S. A. Antonenko, A. I. Sokolov, and K. B. Varnashev,
Phys. Lett. A {\bf 208}, 161 (1995).
\bibitem{8}S. A. Antonenko, A. I. Sokolov, and K. B. Varnashev,
Mol. Mat. {\bf 8}, 175 (1996).
\bibitem{8a}J.-C. Toledano et al., Phys. Rev. B {\bf 31},
7171 (1985).
\bibitem{9}G. Grinstein and D. Mukamel, J. Phys. A: Math. Gen.
{\bf 15}, 233 (1982).
\bibitem{10}K. G. Wilson and M. E. Fisher, Phys. Rev. Lett.
{\bf 28}, 240 (1972).
\bibitem{11}D. Mukamel and S. Krinsky, Phys. Rev. B {\bf 13},
5065 (1976)
\bibitem{12}K. B. Varnashev and A. I. Sokolov, Fizika Tverdogo Tela
(St.Petersburg)
{\bf 38}, 3665 (1996) [Phys. Sol. State {\bf 38}, 1996 (1996)].
\bibitem{13}H. Kawamura, Phys. Rev. B {\bf 38}, 960 (1988);
Phys. Rev. B {\bf 38}, 4916 (1988).
\bibitem{14}A. I. Sokolov, K. B. Varnashev, and A. I. Mudrov,
{\it Critical exponents for the model with unique stable fixed
point from three--loop RG expansions}, report at the International
Conference "Renormalization Group'96" (Dubna, Russia, August 1996),
to appear in the Conference
Proceedings (World Scientific,
Singapore).
\bibitem{15}D. K. De'Bell and D. J. W. Geldart, Phys. Rev. B
{\bf 32}, 4763 (1985).
\bibitem{16}N. A. Shpot, Phys. Lett. A {\bf 133}, 125 (1988);
Phys. Lett. A {\bf 142}, 474 (1989).
\bibitem{17}K. A. Cowley and A. D. Bruce, J. Phys. C {\bf 11},
3577 (1978).
\bibitem{18} J. Sak, Phys. Rev. B {\bf 10}, 3957 (1974).
\bibitem{19}L. S. Goldner and G. Ahlers, Phys. Rev. B {\bf 45},
13129 (1992).
\bibitem{20}J. A. Lipa et. al., Phys. Rev. Lett. {\bf 76},
944 (1996).
\bibitem{21}G. A. Baker, B. G. Nickel, and D. I. Meiron,
Phys. Rev. B {\bf 17}, 1365 (1978).
\bibitem{22}J. C. Le Guillou and J. Zinn--Justin,
Phys. Rev. Lett. {\bf 39}, 95 (1977).
\bibitem{23}D. I. Kazakov, O. V. Tarasov, and A. A. Vladimirov,
Preprint JINR E2--12249 Dubna (1979).
\bibitem{24}G. 't Hooft and M. Veltman, Nucl. Phys. B {\bf 44},
189 (1972).
\bibitem{25}G. 't Hooft, Nucl. Phys. B {\bf 61}, 455 (1973).
\bibitem{26}A. A. Vladimirov, D. I. Kazakov, and O. V. Tarasov,
Zh. Eksp. Teor. Fiz. {\bf 77}, 1035 (1979) [Sov. Phys. JETP
{\bf 50}, 521 (1979)].
\bibitem{27}A. L. Korzhenevskii, Zh. Eksp. Teor. Fiz. {\bf 71},
1434 (1976) [Sov. Phys. JETP {\bf 44}, 751 (1976)].
\bibitem{28}E. Brezin, Le J. C. Guillou, and J. Zinn--Justin,
{\em Phase Transitions and Critical Phenomena} Vol. 6, edited by
C.Domb and M.S.Green, Academic Press, New York, 1976.
\bibitem{29}L. N. Lipatov, Zh. Eksp. Teor. Fiz. {\bf 72},
411 (1977) [Sov. Phys. JETP {\bf 45}, 216 (1977)].
\bibitem{30}E. Brezin, Le J. C. Guillou, and J. Zinn--Justin,
Phys.Rev. D {\bf 15}, 1544 (1977).
\bibitem{31}J. C. Le Guillou and J. Zinn--Justin,
Phys. Rev. B {\bf 21}, 3976 (1980).
\bibitem{32}J. C. Le Guillou and J. Zinn--Justin,
J. Phys. (Paris) Lett. {\bf 46}, L137 (1985).
\bibitem{33}H. Kleinert and S. Thoms,
Phys.Rev. D {\bf 52}, 5926 (1995).
\bibitem{35}B. V. Shabat,
{\em Introduction to the complex analysis} (Nauka, Moscow, 1985).

\end{thebibliography}
\end{document}